\documentclass[runningheads]{llncs}

\usepackage{graphicx}
\usepackage[pagebackref=true,breaklinks=true,colorlinks,bookmarks=false]{hyperref}
\usepackage{amsmath, amssymb}
\usepackage{arydshln}
\usepackage{xcolor}
\usepackage{booktabs}
\usepackage{tabularx}
\usepackage[labelfont=bf, size=small]{caption}

\newcommand{\textBF}[1]{\pdfliteral direct {2 Tr 0.3 w} #1 \pdfliteral direct {0 Tr 0 w}}
\newcommand*\samethanks[1][\value{footnote}]{\footnotemark[#1]}

\begin{document}

\title{Optimization with soft Dice can lead to a volumetric bias}
\titlerunning{Optimization with soft Dice can lead to a volumetric bias}
\author{
    Jeroen Bertels\inst{1}\thanks{J.B. and D.R. have contributed equally to this work.}
    \and David Robben\inst{1, 2}\samethanks
    \and Dirk Vandermeulen\inst{1}
    \and Paul Suetens\inst{1}
}
\authorrunning{J. Bertels and D. Robben et al.}
\institute{
    Processing Speech and Images, ESAT, KU Leuven, Belgium 
    \email{jeroen.bertels@kuleuven.be}
    \and icometrix, Leuven, Belgium
}
\maketitle

\begin{abstract}
Segmentation is a fundamental task in medical image analysis. The clinical interest is often to measure the volume of a structure. To evaluate and compare segmentation methods, the similarity between a segmentation and a predefined ground truth is measured using metrics such as the Dice score. Recent segmentation methods based on convolutional neural networks use a differentiable surrogate of the Dice score, such as soft Dice, explicitly as the loss function during the learning phase. Even though this approach leads to improved Dice scores, we find that, both theoretically and empirically on four medical tasks, it can introduce a volumetric bias for tasks with high inherent uncertainty. As such, this may limit the method's clinical applicability.
\keywords{Segmentation \and Cross-entropy \and Soft Dice \and Volume.}
\end{abstract}

\section{Introduction}
Automatic segmentation of structures is a fundamental task in medical image analysis. Segmentations either serve as an intermediate step in a more elaborate pipeline or as an end goal by itself. The clinical interest often lies in the volume of a certain structure (e.g. the volume of a tumor, the volume of a stroke lesion), which can be derived from its segmentation \cite{Goyal2016}. The segmentation task can also carry inherent uncertainty (e.g. noise, lack of contrast, artifacts, incomplete information).  \\
To evaluate and compare the quality of a segmentation, the similarity between the true segmentation (i.e. the segmentation derived from an expert's delineation of the structure) and the predicted segmentation must be measured. For this purpose, multiple metrics exist. Among others, overlap measures (e.g. Dice score, Jaccard index) and surface distances (e.g. Haussdorf distance, average surface distance) are commonly used \cite{Kamnitsas2017}.\\
The focus on one particular metric, the Dice score, has led to the adoption of a differentiable surrogate loss, the so-called soft Dice \cite{Drozdzal2016,Milletari2016,Sudre2017}, to train convolutional neural networks (CNNs). Many state-of-the-art methods clearly outperform the established cross-entropy losses using soft Dice as loss function \cite{Isensee2018,Bertels2019a}.\\
In this work, we investigate the effect on volume estimation when optimizing a CNN w.r.t. cross-entropy or soft Dice, and relate this to the inherent uncertainty in a task. First, we look into this volumetric bias theoretically, with some numerical examples. We find that the use of soft Dice leads to a systematic under- or overestimation of the predicted volume of a structure, which is dependent on the inherent uncertainty that is present in the task. Second, we empirically validate these results on four medical tasks: two tasks with relatively low inherent uncertainty (i.e. the segmentation of third molars from dental radiographs \cite{DeTobel2017}, BRATS 2018 \cite{brats2018,Menze2015,Bakas2017,Bakas2018}) and two tasks with relatively high inherent uncertainty (i.e. ISLES 2017 \cite{isles2017,Winzeck2018}, ISLES 2018 \cite{isles2018}).\\

\section{Theoretical analysis}
Let us formalize an image into $I$ voxels, each voxel corresponding to a true class label $c_{i}$ with $i=0 \dots I-1$, forming the true class label map $C=[c_{i}]^{I}$. Typical in medical image analysis, is the uncertainty of the true class label map $C$ (e.g. due to intra- and inter-rater variability; see Sect. \ref{sect:uncertainty}). Under the assumption of binary image segmentation with $c_{i} \in \{0,1\}$, a probabilistic label map can be constructed as $Y=[y_{i}]^{I}$, where each $y_{i}=P(c_{i}=1)$ is the probability of $y_{i}$ belonging to the structure of interest. Similarly, we have the maps of voxel-wise label predictions $\hat{C}=[\hat{c}_{i}]^{I}$ and probabilities $\hat{Y}=[\hat{y}_{i}]^{I}$. In this setting, the class label map $\hat{C}$ is constructed from the map of predictions $\hat{Y}$ according to the highest likelihood.\\
The Dice score $\mathcal{D}$ is defined on the label maps as:
\noindent
\begin{equation}
    \mathcal{D}(C, \hat{C}) = \frac{2 |C \cap \hat{C}|}{|C| + |\hat{C}|}
    \label{eq:dice}
\end{equation}
The volumes $\mathcal{V}(C)$ of the true structure and $\mathcal{V}(\hat{C})$ of the predicted structure are then, with $v$ the volume of a single voxel:
\noindent
\begin{equation}
    \mathcal{V}(C) = v\sum_{i=0}^{I-1}c_{i},\ 
    \mathcal{V}(\hat{C}) = v\sum_{i=0}^{I-1}\hat{c}_{i}
\end{equation}\\
In case the label map is probabilistic, we need to work out the expectations: 
\noindent
\begin{equation}
    \mathcal{V}(Y) = v\mathbf{E} [\sum_{i=0}^{I-1}y_{i}],\ 
    \mathcal{V}(\hat{Y}) = v\mathbf{E} [\sum_{i=0}^{I-1}\hat{y}_{i}]
    \label{eq:volume}
\end{equation}\\

\subsection{Risk minimization}
In the setting of supervised and gradient-based training of CNNs \cite{Goodfellow2016} we are performing empirical risk minimization. Assume the CNN, with a certain topology, is parametrized by $\boldsymbol{\theta} \in \Theta$ and represents the functions $\mathcal{H}=\{\mathfrak{h}_{\boldsymbol{\theta}}\}^{|\Theta|}$. Further assume we have access to the entire joint probability distribution $P(\mathbf{x},y)$ at both training and testing time, with $\mathbf{x}$ the information (for CNNs this is typically a centered image patch around the location of $y$) of the network that is used to make a prediction $\hat{y}=\mathfrak{h}_{\boldsymbol{\theta}}(\mathbf{x})$ for $y$. For these conditions, the general risk minimization principle is applicable and states that in order to optimize the performance for a certain non-negative and real-valued loss $\mathcal{L}$ (e.g. the metric or its surrogate loss) at test time, we can optimize the same loss during the learning phase \cite{Vapnik1995}. The risk $\mathcal{R}_{\mathcal{L}}(\mathfrak{h}_{\boldsymbol{\theta}})$ associated with the loss $\mathcal{L}$ and parametrization $\boldsymbol{\theta}$ of the CNN, without regularization, is defined as the expectation of the loss function:
\noindent
\begin{equation}
    \mathcal{R}_{\mathcal{L}}(\mathfrak{h}_{\boldsymbol{\theta}}) = \mathbf{E} [\mathcal{L}(\mathfrak{h}_{\boldsymbol{\theta}}(\mathbf{x}), y)]
    \label{eq:risk}
\end{equation}\\
For years, minimizing the negative log-likelihood has been the gold standard in terms of risk minimization. For this purpose, and due to its elegant mathematical properties, the voxel-wise binary cross-entropy loss ($\mathcal{CE}$) is used: 
\noindent
\begin{equation}
    \mathcal{CE}(\hat{Y},Y) = \sum_{i=0}^{I-1}[\mathcal{CE}(\hat{y}_{i},y_{i})] = -\sum_{i=0}^{I-1}[y_{i}\log\hat{y}_{i} + (1-y_{i})\log(1-\hat{y}_{i})]
    \label{eq:ce}
\end{equation}
More recently, the soft Dice loss ($\mathcal{SD}$) is used in the optimization of CNNs to directly optimize the Dice score at test time \cite{Drozdzal2016,Milletari2016,Sudre2017}. Rewriting Eq. \ref{eq:dice} to its non-negative and real-valued surrogate loss function as in \cite{Drozdzal2016}:
\noindent
\begin{equation}
    \mathcal{SD}(\hat{Y},Y) = 1-\frac{2\sum_{i=0}^{I-1}\hat{y}_{i}y_{i}}{\sum_{i=0}^{I-1}\hat{y}_{i} + \sum_{i=0}^{I-1}y_{i}}
    \label{eq:softdice}
\end{equation}\\

\subsection{Uncertainty}\label{sect:uncertainty}
There is considerable uncertainty in the segmentation of medical images. Images might lack contrast, contain artifacts, be noisy or incomplete regarding the necessary information (e.g. in ISLES 2017 we need to predict the infarction after treatment from images taken before, which is straightforwardly introducing \textit{inherent} uncertainty). Even at the level of the true segmentation, uncertainty exists due to intra- and inter-rater variability. We will investigate what happens with the estimated volume $\mathcal{V}$ of a certain structure in an image under the assumption of having perfect segmentation algorithms (i.e. the prediction is the one that minimizes the empirical risk).\\
Assuming independent voxels, or that we can simplify Eq. \ref{eq:volume} into $J$ independent regions with true uncertainty $p_{j}$ and predicted uncertainty $\hat{p}_{j}$, and corresponding volumes $s_{j}=vn_{j}$, with $n_{j}$ the number of voxels belonging to region $j=0 \dots J-1$ (having each voxel as an independent region when $n_{j}=1$), we get:
\noindent
\begin{equation}
    \mathcal{V}(Y) = \sum_{j=0}^{J-1} (s_{j}p_{j}),\ 
    \mathcal{V}(\hat{Y}) = \sum_{j=0}^{J-1} (s_{j}\hat{p}_{j})
    \label{eq:volumeregions}
\end{equation}\\
We analyze for $\mathcal{CE}$ the predicted uncertainty that minimizes the risk $\mathcal{R}_{\mathcal{CE}}(\mathfrak{h}_{\boldsymbol{\theta}})$:
\noindent
\begin{equation}
    \arg\min_{\hat{Y}}[\mathcal{R}_{\mathcal{CE}}(\mathfrak{h}_{\boldsymbol{\theta}})] = \arg\min_{\hat{Y}}[\mathbf{E} [\mathcal{CE}(\hat{Y}, Y)]]
\end{equation}
We need to find for each independent region $j$:
\noindent
\begin{equation}
    \arg\min_{\hat{p}_{j}}[s_{j}\mathcal{CE}(\hat{p}_j, p_{j})] = \arg\min_{\hat{p}_{j}}[-p_{j}\log\hat{p}_j-(1-p_{j})\log(1-\hat{p}_{j})]
    \label{eq:cemin}
\end{equation}
This function is continuous and its first derivative monotonously increasing in the interval $]0, 1[$. First order conditions w.r.t. $\hat{p}_{j}$ give the optimal value for the predicted uncertainty $\hat{p}_{j}=p_{j}$. With the predicted uncertainty being the true uncertainty, $\mathcal{CE}$ becomes an unbiased volume estimator.\\
We analyze for $SD$ the predicted uncertainty that minimizes the risk $\mathcal{R}_{\mathcal{SD}}(\mathfrak{h}_{\boldsymbol{\theta}})$:
\noindent
\begin{equation}
    \arg\min_{\hat{Y}}[\mathcal{R}_{\mathcal{SD}}(\mathfrak{h}_{\boldsymbol{\theta}})] = \arg\min_{\hat{Y}}[\mathbf{E} [\mathcal{SD}(\hat{Y}, Y)]]
\end{equation}
We need to find for each independent region $j$:
\noindent
\begin{equation}
    \arg\min_{\hat{Y}}[\mathbf{E} [\mathcal{SD}(\hat{Y}, Y)]] = \arg\min_{\hat{p}_{j}}[\mathbf{E}[1-\frac{2\sum_{j=0}^{J-1}s_{j}\hat{p}_{j}p_{j}}{\sum_{j=0}^{J-1}s_{j}\hat{p}_{j} + \sum_{j=0}^{J-1}s_{j}p_{j}}]]
    \label{eq:R_SD}
\end{equation}
This minimization is more complex and we analyze its behavior by inspecting the values of $\mathcal{SD}$ numerically. We will consider the scenarios with only a single region or with multiple independent regions with inherent uncertainty in the image. For each scenario we will vary the inherent uncertainty and the total uncertain volume.\\

\begin{figure}[!b]
    \centering
    \makebox[\linewidth]{
        \includegraphics[width=\textwidth]{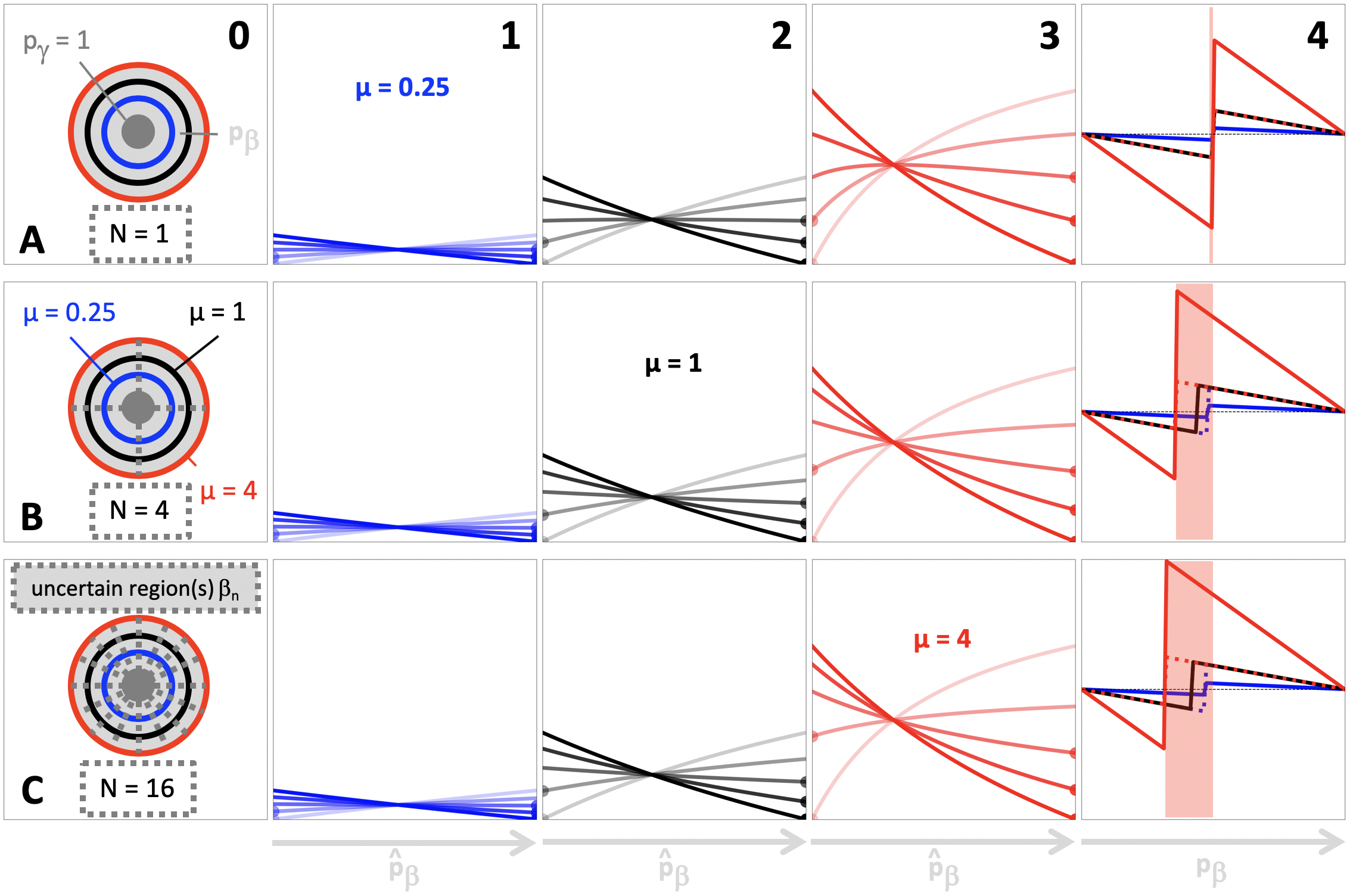}
    }
    \caption{The effects of optimizing w.r.t. $\mathcal{SD}$ for volume ratios: $\mu=0.25$ (blue), $\mu=1$ (black) and $\mu=4$ (red). ROWS \textbf{A}-\textbf{C}: Situations with respectively $N = \{1, 4, 16\}$ independent regions with uncertainty $p_{\beta}$. COLUMN \textbf{0}: Schematic representation of the situation. COLUMNS \textbf{1}-\textbf{3}: $\mathcal{SD}=[0, 1]$ (y-axis) for $p_{\beta}=\{0,0.25,0.5,0.75,1\}$ (respectively with increasing opacity) and $\hat{p}=[0, 1]$ (x-axis). COLUMN \textbf{4}: Influence of $p_{\beta}=[0, 1]$ (x-axis) on volumetric bias (solid lines) or on the error in predicted uncertainty (dashed lines). With the light red area we want to highlight that easier overestimation of the predicted volume occurs due to a higher volume ratio $\mu$ or an increasing number of independent regions $N$.} 
    \label{fig:regions}
\end{figure}

\subsubsection{Single region of uncertainty.}
Imagine the segmentation of an image with $K=3$ independent regions, $\alpha, \beta$ and $\gamma$, as depicted in \textbf{{\small Fig.}} \ref{fig:regions} (\textbf{{\small A0}}). Region $\alpha$ is certainly not part of the structure ($p_{\alpha}=0$, i.e. background), region $\beta$ belongs to the structure with probability $p_{\beta}$ and region $\gamma$ is certainly part of the structure ($p_{\gamma}=1$). Let their volumes be $s_{\alpha}=100$, $s_{\beta}$, $s_{\gamma}=1$, respectively, with $\mu=\frac{s_{\beta}}{s_{\gamma}}=s_{\beta}$ the volume ratio of uncertain to certain part of the structure. Assuming a perfect algorithm, the optimal predictions under the empirical risk from Eq. \ref{eq:R_SD} are:
\noindent
\begin{equation}
    \arg\max_{\hat{p}_{\alpha},\hat{p}_{\beta},\hat{p}_{\gamma}}[\mathbf{E}[\frac{2(s_{\beta}\hat{p}_{\beta}p_{\beta}+s_{\gamma}\hat{p}_{\gamma})}{s_{\alpha}\hat{p}_{\alpha} + s_{\beta}\hat{p}_{\beta} + s_{\gamma}\hat{p}_{\gamma} + s_{\beta}p_{\beta} + s_{\gamma}}]]
\end{equation}
It is trivial to show that $\hat{p}_{\alpha}=0=p_{\alpha}$ and $\hat{p}_{\gamma}=1=p_{\gamma}$ are solutions for this equation. The behavior of $\hat{p}_{\beta}$ w.r.t. $p_{\beta}$ and $\mu$ can be observed qualitatively in \textbf{{\small Fig.}} \ref{fig:regions} (\textbf{{\small A1}}-\textbf{{\small A4}}).
Indeed, only for $p_{\beta}=\{0, 1\}$ the predicted uncertainty $\hat{p}_{\beta}$ is exact. The location of the local minimum in $\hat{p}_{\beta}=[0, 1]$ switches from 0 to 1 when $p_{\beta}=0.5$. Therefore, when $p_{\beta}$ decreases or increases from 0.5 (different opacity in \textbf{{\small A1}}-\textbf{{\small A3}}), respectively under- or overestimation will occur (\textbf{{\small A4}}). The resulting volumetric bias will be highest when the inherent uncertainty $p_{\beta}=0.5$ and decreases towards the points of complete certainty, being always 0 or 1. The effect of the volume ratio $\mu$ (colors) is two-fold. With $\mu$ increasing, the optimal loss value increases (\textbf{{\small A1}}-\textbf{{\small A3}}) and the volumetric bias increases (\textbf{{\small A4}}; solid lines). However, the error on the estimated uncertainty is not influenced by $\mu$ (\textbf{{\small A4}}; dashed lines).\\

\subsubsection{Multiple regions of uncertainty.}
In a similar way we can imagine the segmentation of a structure with $K=N+2$ independent regions, for which we further divided the region $\beta$ into $N$ equally large independent sub-regions $\beta_{n}$ with $n=0 \dots N-1$. Let us further assume they have the same inherent uncertainty $p_{\beta_{n}}=p_{\beta}$ and volume ratio $\mu_{\beta_{n}}=\frac{\mu_{\beta}}{N}$ (in order to keep the total uncertain volume the same). If we limit the analysis to a qualitative observation of \textbf{{\small Fig.}} \ref{fig:regions} with $N=4$ (\textbf{{\small B0}}-\textbf{{\small B4}}) and $N=16$ (\textbf{{\small C0}}-\textbf{{\small C4}}), we notice three things. First, the uncertainty $p_{\beta}$ for which under- or overestimation will happen decreases (\textbf{{\small A4}}, \textbf{{\small B4}}, \textbf{{\small C4}}). Second, this effect is proportional with $\mu$ and the maximal error on the predicted uncertainty becomes higher (\textbf{{\small B0}}-\textbf{{\small B4}}, \textbf{{\small C0}}-\textbf{{\small C4}}). Third, there is a trend towards easier volumetric overestimation and with the maximal error being more pronounced when the number of regions increases (\textbf{{\small A4}}, \textbf{{\small B4}}, \textbf{{\small C4}}).\\

\section{Empirical analysis}
In this section we will investigate whether the aforementioned characteristics can be observed under real circumstances. In a practical scenario, the joint probability distribution $P(\mathbf{x},y)$ is unknown and presents itself as a training set. The risk $\mathcal{R}_{\mathcal{L}}$ (Eq. \ref{eq:risk}) becomes empirical, where the expectation of the loss function becomes the mean of the losses across the training set. Furthermore, the loss $\mathcal{L}$ absorbs the explicit (e.g. weight decay, L2) or implicit (e.g. early stopping, dropout) regularization, which is often present in some aspect of the optimization of CNNs. Finally, the classifier is no longer perfect and additionally to the inherent uncertainty in the task we now have inherent uncertainty introduced by the classifier itself.\\
To investigate how these factors impact our theoretical findings, we train three models with increasing complexity: LR (logistic regression on the input features), ConvNet (simpler version of the next) and U-Net. We use five-fold cross-validation on the training images from two tasks with relatively low inherent uncertainty (i.e. lower-left third molar segmentation from panoramic dental radiographs (MOLARS) \cite{DeTobel2017}, BRATS 2018 \cite{brats2018}) and from two tasks with relatively high inherent uncertainty (i.e. ISLES 2017 \cite{isles2017}, ISLES 2018 \cite{isles2018}). Next, we describe the experimental setup, followed by a dissemination of the predicted volume errors $\Delta \mathcal{V}(\hat{Y}, Y)=\mathcal{V}(\hat{Y})-\mathcal{V}(Y)$ by $\mathcal{CE}$ and $\mathcal{SD}$ trained models.\\

\subsection{Task description and training}
We (re-)formulate a binary segmentation task for each dataset having one (multi-modal) input, and giving one binary segmentation map as output (for BRATS 2018 we limit the task to whole tumor segmentation). For the 3D public benchmarks we use all of the provided images, resampled to an isotropic voxel-size of 2 mm, as input (for both ISLES challenges we omit perfusion images). In MOLARS (2D dataset from \cite{DeTobel2017}), we first extract a 448x448 ROI around the geometrical center of the lower-left third molar from the panoramic dental radiograph. We further downsample the ROI by a factor of two. The output is the segmentation of the third molar, as provided by the experts. All images are normalized according to the dataset's mean and standard deviation.\\
For our U-Net model we start from the successful No New-Net implementation during last year's BRATS challenge \cite{Isensee2018}. We adapt it with three 3x3(x3) average pooling layers with corresponding linear up-sampling layers and strip the instance normalization layers. Each level has two 3x3(x3) convolutional layers before and after the pooling and up-sampling layer, respectively, with [[10, 20], [20, 10]], [[20, 40], [40, 20]], [[40, 80], [80, 40]] and [40, 20] filters. For the ConvNet model, we remove the final two levels. The LR model uses the inputs directly for classification, thus performing logistic regression on the input features.\\
The images are augmented intensively during training and inputs are central image crops of 162x162x108 (in MOLARS 243x243). We train the models w.r.t. $\mathcal{CE}$ or $\mathcal{SD}$ with ADAM, without any explicit regularization, and with the initial learning rate set at $10^{-3}$ (for LR model at 1). We lower the learning rate by a factor of five when the validation loss did not improve over the last 75 epochs and stop training with no improvement over the last 150 epochs.\\

\subsection{Results and discussion}
\begin{table}[!b]
    \centering
    \caption{Empirical results for cross-entropy ($\mathcal{CE}$), soft Dice score ($1-\mathcal{SD}$) and volume error ($\Delta \mathcal{V}$; in $10^{2}$ \textit{pixels} or $ml$) metrics for models optimized w.r.t. $\mathcal{CE}$ and $\mathcal{SD}$ losses. Significant volumetric underestimations in \textit{italic} and overestimations in \textbf{bold}.}
    \begin{tabular*}{\linewidth}{@{\extracolsep{\fill}}l|l|rr|rr|rr}
        \toprule
                  & \multicolumn{1}{r|}{Model $\rightarrow$} & \multicolumn{2}{c|}{LR} & \multicolumn{2}{c|}{ConvNet} & \multicolumn{2}{c}{U-Net} \\
                  & \multicolumn{1}{r|}{Training loss $\rightarrow$} &     \multicolumn{1}{c}{$\mathcal{CE}$} &       \multicolumn{1}{c|}{$\mathcal{SD}$} &    \multicolumn{1}{c}{$\mathcal{CE}$} &      \multicolumn{1}{c|}{$\mathcal{SD}$} &    \multicolumn{1}{c}{$\mathcal{CE}$} &     \multicolumn{1}{c}{$\mathcal{SD}$} \\
        Dataset $\downarrow$   & Metric $\downarrow$ &          &            &         &           &         &          \\
        \midrule
        MOLARS (2D)     & $\mathcal{CE}(\hat{Y}, Y)$   &   0.240 &     5.534 &  0.194 &    1.456   &  0.024 &   0.103 \\
                  & $1-\mathcal{SD}(\hat{Y}, Y)$   &   0.068 &     0.153 &  0.150 &    0.270   &  0.865 &   0.931 \\
                  \\ [-1.02em] \cdashline{2-8} \\ [-1.02em]
                  & $\Delta \mathcal{V}(\hat{Y}, Y)$ ($10^{2}$ \textit{pixels})  &  -0.069 &  \textBF{302.3}  & -0.276 &  \textBF{87.09}   &  0.092 &  -0.187 \\
        \midrule  
        BRATS 2018 (3D) & $\mathcal{CE}(\hat{Y}, Y)$   &   0.039 &     0.173 &  0.030 &    0.069   &  0.012 &   0.027 \\
                  & $1-\mathcal{SD}(\hat{Y}, Y)$   &   0.080 &     0.355 &  0.196 &    0.715   &  0.585 &   0.820 \\
                  \\ [-1.02em] \cdashline{2-8} \\ [-1.02em]
                  & $\Delta \mathcal{V}(\hat{Y}, Y)$ ($ml$)  &  -2.841 &   \textBF{276.4}   &  3.936 &   \textBF{19.93}    & \textit{-6.778} &  -1.905 \\
        \midrule
        ISLES 2017 (3D) & $\mathcal{CE}(\hat{Y}, Y)$   &   0.025 &     0.155 &  0.018 &    0.069   &  0.014 &   0.066 \\
                  & $1-\mathcal{SD}(\hat{Y}, Y)$   &   0.099 &     0.255 &  0.114 &    0.321   &  0.188 &   0.340 \\
                  \\ [-1.02em] \cdashline{2-8} \\ [-1.02em]
                  & $\Delta \mathcal{V}(\hat{Y}, Y)$ ($ml$)  &  15.71  &    \textBF{82.42}  & -4.227 &   \textBF{23.83}    & -2.875 &  \textBF{13.44} \\
        \midrule
        ISLES 2018 (3D) & $\mathcal{CE}(\hat{Y}, Y)$   &   0.055 &     0.225 &  0.044 &    0.139   &  0.029 &   0.128 \\
                  & $1-\mathcal{SD}(\hat{Y}, Y)$   &   0.136 &     0.329 &  0.200 &    0.449   &  0.362 &   0.518 \\
                  \\ [-1.02em] \cdashline{2-8} \\ [-1.02em]
                  & $\Delta \mathcal{V}(\hat{Y}, Y)$ ($ml$)  &   0.773 &    \textBF{34.03}  & -0.374 &   \textBF{12.44}    & -0.878 &   \textBF{5.442} \\
        \bottomrule
    \end{tabular*}
    \label{tab:results}
\end{table}
In \textbf{{\small Table.}} \ref{tab:results} the results are shown for each dataset (i.e. MOLARS, BRATS 2018, ISLES 2017, ISLES 2018), for each model (i.e. LR, ConvNet, U-Net) and for each loss (i.e. $\mathcal{CE}$, $\mathcal{SD}$) after five-fold cross-validation. We performed a pairwise non-parametric significance test (bootstrapping) with a p-value of 0.05 to assess inferiority or superiority between pairs of optimization methods.\\
Optimizing the $\mathcal{CE}$ loss reaches significantly higher log-likelihoods under all circumstances, while soft Dice \textit{scores} (i.e. $1-\mathcal{SD}$) are significantly higher for $\mathcal{SD}$ optimized models. Looking at the volume errors $\Delta \mathcal{V}(\hat{Y}, Y)$, the expected outcomes are, more or less, confirmed. For the LR and ConvNet models, $\mathcal{CE}$ optimized models are unbiased w.r.t. volume estimation. For these models, $\mathcal{SD}$ optimization leads to significant overestimation due to the remaining uncertainty, partly being introduced by the models themselves.\\
The transition to the more complex U-Net model brings forward two interesting observations. First, for the two tasks with relatively low inherent uncertainty (i.e. MOLARS, BRATS 2018), the model is able to reduce the uncertainty to such an extent it can avoid significant bias on the estimated volumes. The significant underestimation for $\mathcal{CE}$ in BRATS 2018 can be due to the optimization difficulties that arise in circumstances with high class-imbalance. Second, although the model now has the ability to extend its view wide enough and propagate the information in a complex manner, the inherent uncertainty that is present in both of the ISLES tasks, brings again forward the discussed bias. In ISLES 2017, having to predict the infarction after treatment straightforwardly introduces uncertainty. In ISLES 2018, the task was to detect the acute lesion, as observed on MR DWI, from CT perfusion-derived parameter maps. It is still unknown to what extent these parameter maps contain the necessary information to predict the lesion.\\
The $\mathcal{CE}$ optimized U-Net models result in Dice scores (Eq. \ref{eq:dice}) of 0.924, 0.763, 0.177 and 0.454 for MOLARS, BRATS 2018, ISLES 2017 and ISLES 2018, respectively. The Dice scores obtained with their $\mathcal{SD}$ optimized counterparts are significantly higher, respectively 0.932, 0.826, 0.343 and 0.527. This is in line with recent theory and practice from \cite{Bertels2019a} and justifies $\mathcal{SD}$ optimization when the segmentation quality is measured in terms of Dice score.\\

\section{Conclusion}
It is clear that, in cases with high inherent uncertainty, the estimated volumes with soft Dice-optimized models are biased, while cross-entropy-optimized models predict unbiased volume estimates. For tasks with low inherent uncertainty, one can still favor soft Dice optimization due to a higher Dice score.\\
We want to highlight the importance of choosing an appropriate loss function w.r.t. the goal. In a clinical setting where volume estimates are important and for tasks with high or unknown inherent uncertainty, optimization with cross-entropy can be preferred.\\

\subsubsection{Acknowledgements.}
J.B. is part of NEXIS \cite{nexis}, a project that has received funding from the European Union's Horizon 2020 Research and Innovations Programme (Grant Agreement \#780026). D.R. is supported by an innovation mandate of Flanders Innovation and Entrepreneurship (VLAIO).

\bibliographystyle{splncs04}
\bibliography{bibliography}

\end{document}